\documentclass[aps,prd,superscriptaddress,showpacs,preprint]{revtex4}
\usepackage{graphicx, bm}
\begin{document}
\draft
\title{Constraints on Left-Right Symmetric and $E_6$ Superstring
Models from the Number of Light Neutrino Species}

\author{ A. Guti\'errez-Rodr\'{\i}guez}
\affiliation{\small Facultad de F\'{\i}sica, Universidad Aut\'onoma de Zacatecas\\
         Apartado Postal C-580, 98060 Zacatecas, Zacatecas M\'exico.\\
         Cuerpo Acad\'emico de Part\'{\i}culas Campos y Astrof\'{\i}sica.}
\author{M. A. Hern\'andez-Ru\'{\i}z}
\affiliation{\small Facultad de Ciencias Qu\'{\i}micas, Universidad Aut\'onoma de Zacatecas\\
        Apartado Postal 585, 98060 Zacatecas, Zacatecas M\'exico.\\}
\author{M. A. P\'erez}
\affiliation{\small Departamento de F\'{\i}sica, CINVESTAV.\\
             Apartado Postal 14-740, 07000, M\'exico D.F., M\'exico.}
\author{F. P\'erez-Vargas $^1$}

\date{\today}

\begin{abstract}

Using the experimental value for the rate $R^{LEP}
_{exp}=\Gamma_{inv}/\Gamma_{l\bar l}=5.942\pm 0.016$, we derive
constraints on the number of light neutrino species with the
invisible width method in the framework of a left-right symmetric
model (LRSM) and a $E_6$ superstring model as a function of their
respective mixing angles for the neutral vector gauge bosons.
Using the above LEP result we get the constraints $-1.6\times
10^{-3}\leq\phi \leq 1.1\times 10^{-3}$ and $-1.3\times
10^{-2}\leq\phi_{E_6} \leq 0.4\times 10^{-2}$, which are stronger
than those obtained in previous studies of these models.
\end{abstract}

\pacs{14.60.Lm,12.15.Mm, 12.60.-i
\vspace*{2cm}\noindent  E-mail: $^{1}$alexgu@planck.reduaz.mx,
$^{2}$mahernan@uaz.edu.mx, $^{3}$mperez@fis.cinvestav.mx}

\vspace{5mm}

\maketitle


\section{Introduction}

The number of fermion generations, which is associated to the
number of light neutrinos, is one of the most important
predictions of the Standard Model of the electroweak interactions
(SM) \cite{Glashow,Weinberg,Salam}. In the SM the decay width of
the $Z_1$ boson into each neutrino family is calculated to be
$\Gamma_{\nu\bar \nu}= 166.3\pm 1.5$ $MeV$ \cite{Data06}.
Additional generations, or other new weakly interacting particles
with masses below $M_{Z_1}/2$, would lead to a decay width of the
$Z_1$ into invisible channels larger than the SM prediction for
three families while a smaller value could be produced, for
example, by the presence of one or more right-handed neutrinos
mixed with the left-handed ones \cite{Jarlskog}. Thus the number
of light neutrino generations $N_\nu$, defined as the ratio
between the measured invisible decay width of the $Z_1$,
$\Gamma_{inv}$, and the SM expectation $\Gamma_{\nu\bar \nu}$ for
each neutrino family, need not be an integer number and has to be
measured with the highest possible accuracy.

The most precise measurement of the number of light $(m_\nu \ll 1
\hspace{1mm} GeV)$ active neutrino types, and therefore the number
of associated fermion families, comes from the invisible $Z_1$
width $\Gamma_{inv}$, obtained by subtracting the observed width
into quarks and charged leptons from the total width obtained from
the lineshape. The number of effective neutrinos $N_\nu$ is given
by \cite{M.Carena}

\begin{equation}
N_\nu =\frac{\Gamma_{inv}}{\Gamma_{l\bar l}}(\frac{\Gamma_{l\bar
l}}{\Gamma_{\nu \bar \nu}})_{SM},
\end{equation}

\noindent where $(\frac{\Gamma_{l\bar l}}{\Gamma_{\nu \bar
\nu}})_{SM}$, the SM expression for the ratio of widths into a
charged lepton and a single active neutrino, is introduced to
reduce the model dependence. The experimental value from the four
LEP experiments is $N_\nu$= 2.9841$ \pm $0.0083
\cite{Data06,Abbaneo,Acciarri,Buskulic}, excluding the possibility
of a fourth family unless the neutrino is very heavy. This result
is in agreement with cosmological constraints on the number of
relativistic species around the time of Big Bang nucleosynthesis,
which seems to indicate the existence of three very light neutrino
species \cite{V.Barger,Hannestad,Cyburt,Kasuhide}. On the other
hand, the LEP result is about two sigmas away from the SM
expectation, $N_\nu$= 3. While not statistically significant, this
result suggests that the $Z\nu\bar\nu$-couplings might be
suppressed with respect to the SM value
\cite{M.Carena,M.Maya,Huerta}.

Using the experimental value for
$R^{LEP}_{exp}=\frac{\Gamma_{inv}}{\Gamma_{l\bar l}}=5.942\pm
0.016$ \cite{Abbaneo,Acciarri,Buskulic}, we will determine the
allowed region for $N_\nu$ as a function of $\phi$ and
$\phi_{E_6}$, the mixing angles between the light and the heavy
neutral gauge bosons $Z_1$ and $Z_2$ of the LRSM
\cite{A.Gutierrez,Gutierrez01,Gutierrez02,Gutierrez03,Gutierrez04}
and $E_6$ superstring models
\cite{London,Capstick,Leike,Alam,Langacker,Aytekin}. In this way
we will be able to get bounds on these mixing angles using the LEP
data on the number of light neutrino species. On the other hand,
if one assumes that the results for $\Gamma_{inv}$ and
$\Gamma_{l\bar l}$ for a future TESLA-like Giga-$Z_1$ experiment
agree with the central values obtained at LEP, one would get
$(\frac{\Gamma_{inv}}{\Gamma_{l\bar l}})^{Giga-Z_1}=5.942\pm
0.012$ (most conservative) or $(\frac{\Gamma_{inv}}{\Gamma_{l\bar
l}})^{Giga-Z_1}=5.942\pm 0.006$ (most optimistic) \cite{M.Carena}.
In this case we estimate also a limit for these mixing angles.

This paper is organized as follows: In Sect. II we present the
numerical computation and, finally, we summarize our results in
Sec. III.

\section{Results}

We will take advantage of our previous work on the LRSM
\cite{A.Gutierrez,Gutierrez01,Gutierrez02,Gutierrez03} and the
$E_6$ superstring model \cite{A.Gutierrez1,A.Gutierrez2} and we
will calculate the partial widths for $Z_1 \to l\bar l$ and $Z_1
\to \nu \bar \nu $ using the transition amplitudes given in Eqs.
(21) and (22) of Ref.
\cite{A.Gutierrez,Gutierrez01,Gutierrez02,Gutierrez03} and the
couplings of the two neutral gauge bosons involved in the $E_6$
model to fermions given in Refs. \cite{Capstick, Aytekin}. In
order to compare the respective expressions with the experimental
result for the number of light neutrinos species $N_\nu$, we will
use the definition for $N_\nu$ in a SM analysis \cite{M.Acciarri}
given in Eq. (1) and the LEP result for the $Z_1$ invisible width
\cite{Data06,Abbaneo,Acciarri,Buskulic},

\begin{equation}
R^{LEP}_{exp}=(\frac{\Gamma_{inv}}{\Gamma_{l\bar l}})=5.942\pm
0.016.
\end{equation}

This definition replaces the expression
$N_\nu=\frac{\Gamma_{inv}}{\Gamma_{\nu \bar \nu}}$ since (2)
reduces the influence of the top quarks mass. To get information
on
the meaning of $N_\nu$ in the LRSM model we should define the
corresponding expression \cite{M.Maya,Huerta},

\begin{equation}
(N_{\nu})_{LRSM}=R_{exp}(\frac{\Gamma_{l\bar l}}{\Gamma_{\nu
\bar\nu}})_{_{LRSM}}.
\end{equation}

This new expression is a function of the mixing angle $\phi$, and
in this case the quantity defined as the number of light neutrinos
species is not a constant and not necessarily an integer. Also,
$(N_\nu)_{LRSM}$ in formula (3) is independent from the $Z_2$ mass
and therefore depends only of the mixing angle $\phi$ of the LRSM.
Experimental values for $\Gamma_{inv}$ and for $\Gamma_{l\bar l}$
are reported in literature which, in our case, can give a bound
for the angle $\phi$. However, we can look to those experimental
numbers in another way. The partial widths $\Gamma_{inv}=499.0\pm
1.5$ $MeV$ and $\Gamma_{l\bar l}=83.984\pm 0.086$ $MeV$ were
reported recently \cite{Data06}, but we use the value given by (3)
for the $R_{exp}$ rate of Ref. \cite{Abbaneo,Acciarri,Buskulic}.
All these measurements are independent of any model and can be
fitted with the LRSM parameter $(N_\nu)_{LRSM}$ in terms of
$\phi$.

The same argument applies to the case of the $E_6$ superstring
model and we will be able to get also a limit on its respective
mixing angle $\phi_{E_6}$ between the light $Z_1$ and heavy
$Z_\theta$ gauge bosons involved in this model
\cite{London,Capstick,Leike,Alam,Langacker,Aytekin,A.Gutierrez2}.

In order to estimate a limit for the number of light neutrino
species $(N_\nu)_{_{LRSM}}$ in the framework of a left-right
symmetric model, we plot the expression (3) in Fig. 1. In this
figure we show the allowed region for $(N_\nu)_{_{LRSM}}$ as a
function of $\phi$ with $90\%$ C.L. The allowed region is the
inclined band that is a result of both factors in Eq. (3). In this
figure $(\frac{\Gamma_{l\bar l}}{\Gamma_{\nu \bar\nu}})_{_{LRSM}}$
gives the inclination while $R_{exp}$ gives the broading. This
analysis was done using the experimental value given in Eq. (2)
for $R_{exp}$ reported by \cite{Abbaneo,Acciarri,Buskulic} with a
$90\%$ C.L. In the same figure we show the SM $(\phi=0)$ result at
$90\%$ C.L. with the dashed horizontal lines.  The allowed region
in the LRSM (dotted line) for $(N_\nu)_{_{LRSM}}$ is wider that
the one for the SM.

If now we reverse the arguments, that is to say we fix the number
of neutrinos in the LRSM to be three then the theoretical
expression for $R$ will be given by

\begin{equation}
R_{LRSM}=\frac{3\Gamma_{\nu\bar\nu}}{\Gamma_{l\bar l}}.
\end{equation}

The plot of this quantity as function of the mixing angle $\phi$
is shows in Fig. 2. The horizontal lines give the experimental
region at $90 \%$ C.L. From the figure we observe that the
constraint for the angle $\phi$ is:

\begin{equation}
-1.6\times 10^{-3}\leq\phi \leq 1.1\times 10^{-3},
\end{equation}

\noindent which is about one order of magnitude stronger than the
one obtained in previous studies of the LRSM
\cite{M.Maya,Huerta,J.Polak,Adriani} and it is consistent with our
results obtained recently with the LEP data obtained for the
process $e^+e^- \to \tau^+\tau^-\gamma$ \cite{A.Gutierrez2}.

In the case of a future TESLA-like Giga-$Z_1$ experiment
\cite{M.Carena} we obtain the following bounds for the mixing
angle $\phi$:

\begin{eqnarray}
-1.1\times 10^{-3}&\leq &\phi\leq 0.9 \times 10^{-3}, \hspace{2mm} \mbox{(most conservative)},\\
-0.8\times 10^{-3}&\leq &\phi\leq 0.33 \times 10^{-3},
\hspace{2mm} \mbox{(most optimistic)}.
\end{eqnarray}

Finally, in Figs. 3 and 4 we present the respective plots for
$(N_\nu)_{E_6}$ and $R_{E_6}$ as a function of $\phi_{E_6}$. In
this case, we get the following constraints on this mixing angle,

\begin{equation}
-1.3\times 10^{-2}\leq\phi_{E_6} \leq 0.4\times 10^{-2},
\end{equation}

\noindent if use is made of the LEP results and

\begin{eqnarray}
-1.1\times 10^{-2}&\leq & \phi_{E_6}\leq 0.2\times
10^{-2} \hspace{2mm} \mbox{(most conservative),}\\
-0.75\times 10^{-2}&\leq & \phi_{E_6}\leq 0.1\times 10^{-2}
\hspace{2mm} \mbox{(most optimistic)},
\end{eqnarray}

\noindent for the case of the expected parameters of the future
TESLA-like Giga-$Z_1$ experiment.

The constraints obtain in the Eqs. (8), (9) and (10) for the
mixing angle $\phi_{E_6}$ are about one order of magnitude
stronger than the one obtained in previous studies of the $E_6$
superstring models
\cite{Barger,Barger1,Hewett,Hewett1,Langacker1}.

\section{Conclusions}

We have determined a bound on the number of light neutrinos
species in the frame work of a left-right symmetric model and a
$E_6$ superstring model as a function of the mixing angles $\phi$
and $\phi_{E_6}$. Using these results and the LEP values obtained
for $N_\nu$, we were able to put limits on the LRSM and $E_6$
superstring mixing angles $\phi$ and $\phi_{E_6}$ which are
stronger than those obtained in previous studies of these models.
If new precision measurements find small deviations from three for
$N_\nu$, these models may explain these deviations with small
values of their mixing angles $\phi$ and $\phi_{E_6}$. We have
found that the new data on
$R_{exp}=\frac{\Gamma_{inv}}{\Gamma_{l\bar l}}$ can shrink
appreciably the allowed regions for $N_\nu$ in the LRSM and the
$E_6$ superstring model. As expected, in the limit of vanishing
$\phi$, $\phi_{E_6}$ and $M_{Z_\theta}\to \infty$ we recover the
bound on $N_\nu$ for the SM previously obtained in the literature
\cite{Data06,Abbaneo,Acciarri,Buskulic,M.Carena}.

\vspace{2cm}

\newpage

\begin{center}
{\bf Acknowledgments}
\end{center}

We would like to thank O. G. Miranda for useful discussions. This work was supported
by CONACyT and SNI (M\'exico).

\newpage

\begin{figure}[t]
\centerline{\scalebox{0.85}{\includegraphics{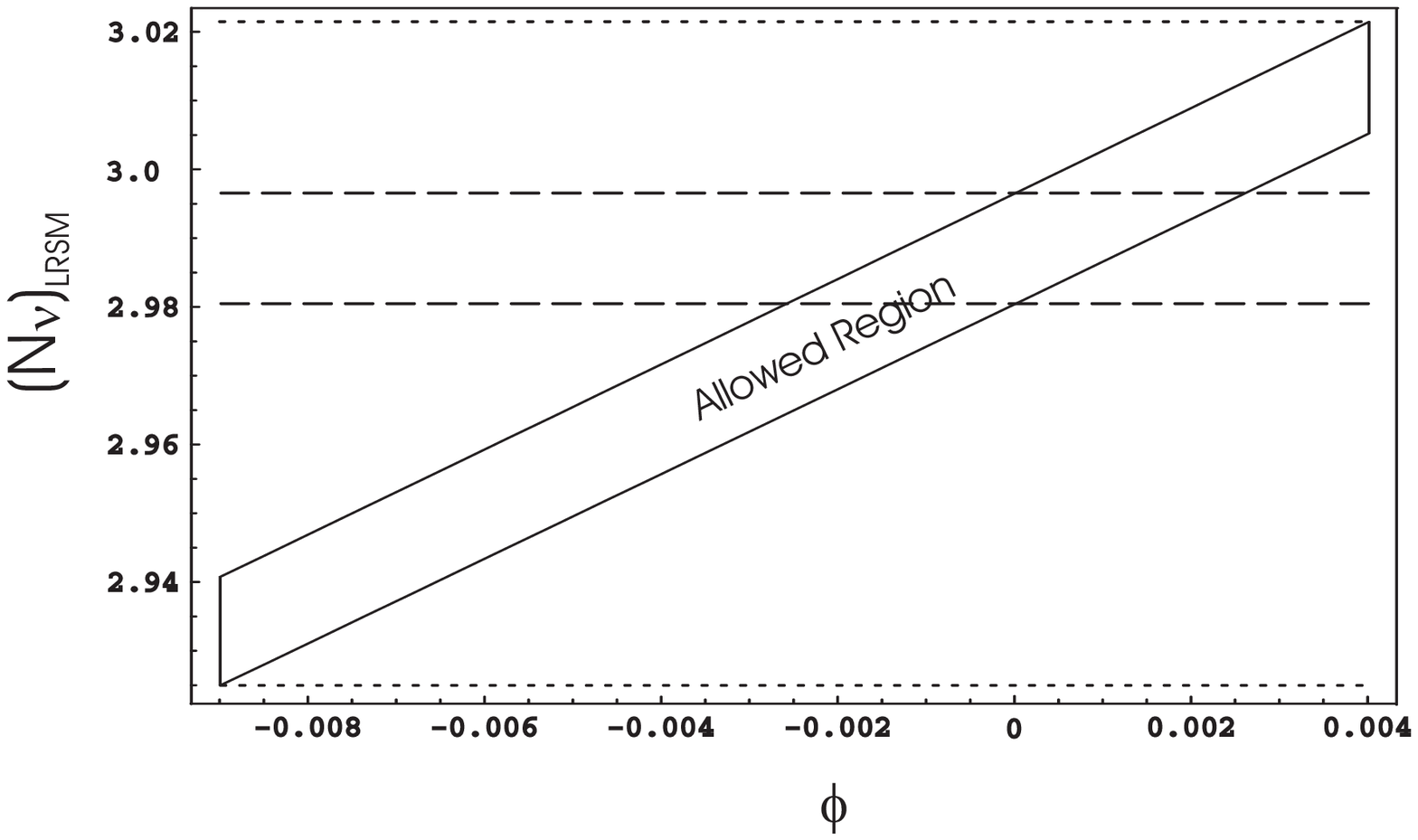}}}
\caption{ \label{fig:gamma} Allowed region for $(N_\nu)_{LRSM}$ as
a function of the mixing angle $\phi$ with the experimental value
$R^{LEP}_{exp}$. The dashed lines shows the SM allowed region for
$N_\nu$ at $90\%$ C.L., while the dotted lines shows the same
result for the LRSM.}
\end{figure}

\begin{figure}[t]
\centerline{\scalebox{0.87}{\includegraphics{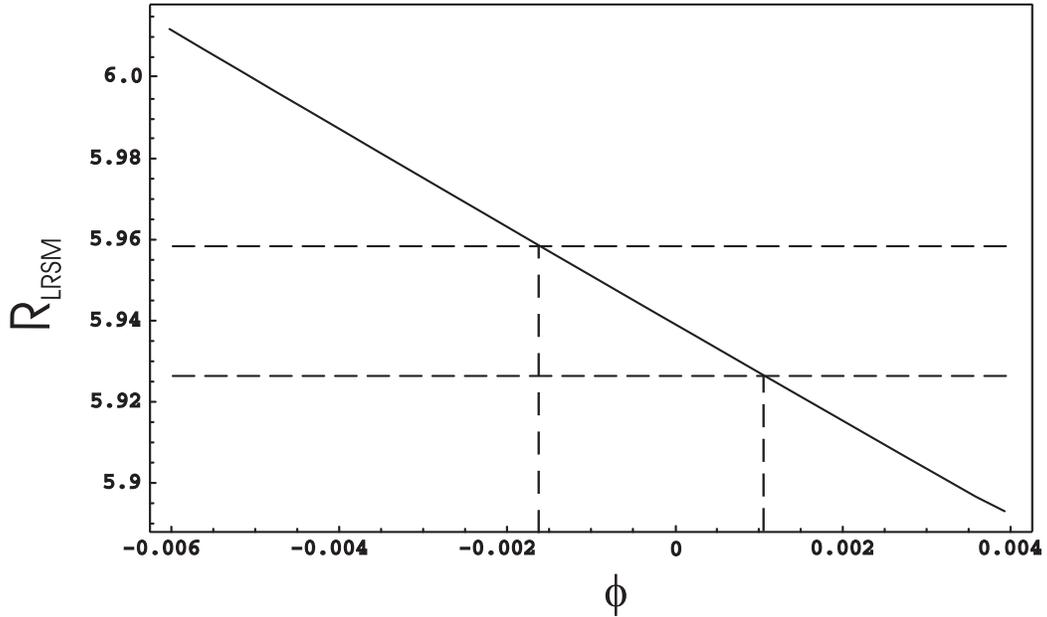}}}
\caption{ \label{fig:gamma} The curve shows the shape for
$R_{LRSM}$ as a function of the mixing angle $\phi$. The dashed
lines shows the experimental region for $R^{LEP}_{exp}$ at $90\%$
C.L. and the dashed vertical region indicates the allowed region
for the mixing angle.}
\end{figure}

\begin{figure}[t]
\centerline{\scalebox{0.9}{\includegraphics{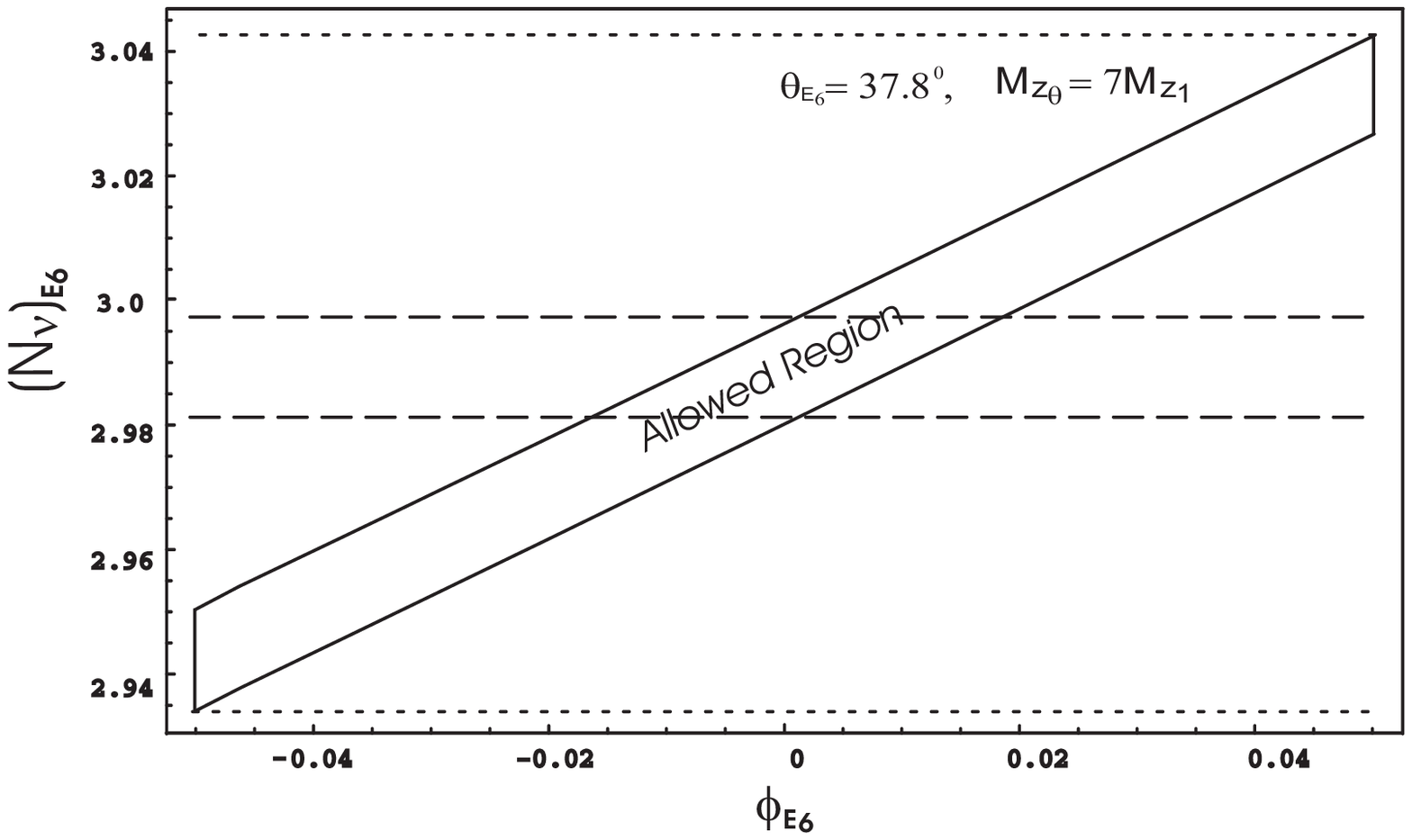}}}
\caption{ \label{fig:gamma} The same as in Fig. 1, but for $E_6$,
with $\theta_{E_6}=37.8^o$ and $M_{Z_\theta}=7M_{Z_1}$.}
\end{figure}

\begin{figure}[t]
\centerline{\scalebox{0.9}{\includegraphics{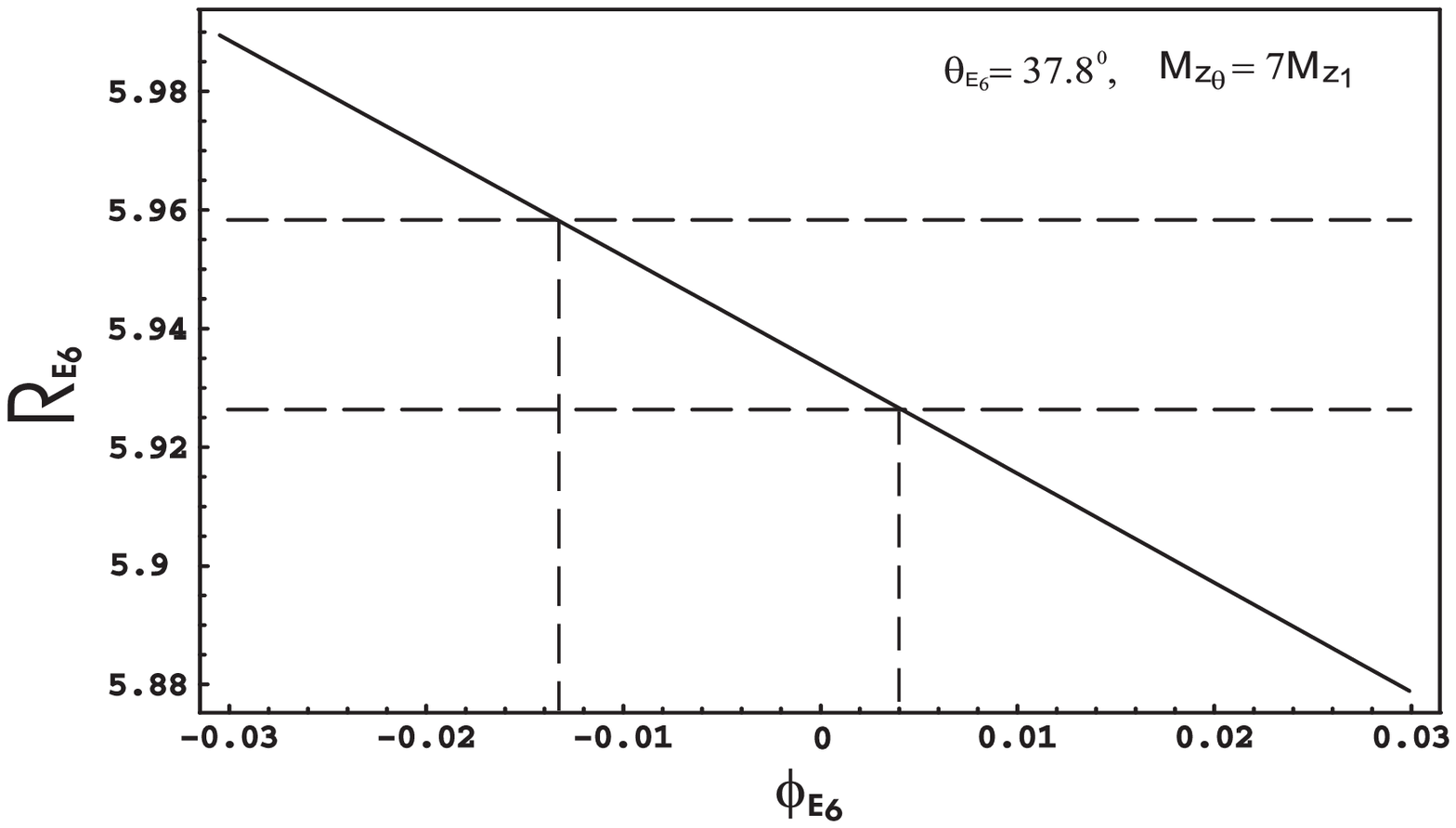}}}
\caption{ \label{fig:gamma} The same as in Fig. 2, but for $E_6$,
with $\theta_{E_6}=37.8^o$ and $M_{Z_\theta}=7M_{Z_1}$.}
\end{figure}

\end{document}